\documentclass[5p,numbered]{elsarticle}
\usepackage{graphicx}

\title{Probing the Ground State Properties of Iron-based Superconducting Pnictides and Related Systems by Muon-Spin Spectroscopy}

\author[psi]{A.~Amato\corref{cor1}}
\author[psi]{R.~Khasanov}
\author[psi]{H.~Luetkens}
\author[dresden]{H.-H.~Klauss}
\cortext[cor1]{Corresponding author}
\address[psi]{Laboratory for Muon-Spin Spectroscopy, Paul Scherrer Institut, CH-5232 Villigen PSI, Switzerland}
\address[dresden]{Institut f\"ur Festk\"orperphysik, TU Dresden, D-01069 Dresden, Germany}
\begin{document}
\sloppy
\begin{abstract}
In this short review, we attempt to give a comprehensive discussion of studies performed to date by muon-spin spectroscopy (more precisely the relaxation and rotation technique, also know as $\mu$SR) on the recently discovered layered iron-based superconductors. On one side, $\mu$SR has been used to characterized the magnetic state of different families of layered iron-based systems. Similarly the subtle interplay of the magnetic state and the structural transition present in some families has been investigated. We will also discuss the information provided by this technique on the interaction between the magnetic state and the superconducting phase. Finally the $\mu$SR technique has been used to investigate the magnetic penetration depth of the superconducting ground state. The study of its absolute value, temperature and magnetic field dependence provides crucial tests for investigating possible unconventional  superconducting states in such systems. 
\end{abstract}
\maketitle
\section{Introduction}
The recent discovery of superconductivity (SC) in layered iron pnictides \cite{kamihara} has triggered 
a surge of research activity devoted to understanding their magnetic and superconducting states, 
as well as to disentangling their fundamental interplay. Probably the foremost motivation 
of such interest is based on the apparent striking similarity between such systems and the well-studied 
superconducting cuprates. Hence, similar to the cuprates, SC takes place mainly in crystal layers (in this case FeAs) with the rest of the structure acting as a charge reservoir. Moreover, a remarkable parallel with the cuprates can be drawn from the observation that SC appears when doping away from an antiferromagnetically ordered mother compound, suggesting the importance of magnetic fluctuations in the mechanism of Cooper pairs formation.

For the research on cuprates, the muon-spin spectros\-copy ($\mu$SR) technique has played a key role, on one side by investigating the interplay between magnetism and superconductivity, and on the other side by providing fundamental results about the nature of the superconducting state. It is therefore not surprising that this technique has been, at a very early stage, widely involved in the study of the layered iron pnictides. The aim of this article is to provide a short review of the $\mu$SR studies available to date and to discuss the main insights provided by these measurements.

As developed in the next Section, the $\mu$SR technique possesses few features which might be exploited to provide key knowledge when studying the iron-based layered pnictides. Its very high sensitivity to internal magnetic fields allows one to investigate weak magnetic states and to determine very precisely the temperature dependence of the magnetic order parameter. As being a local probe technique, $\mu$SR is widely recognized as one of the key experiment to test for possible microscopic coexistence between different ground states. In view of the vicinity of the magnetic and superconducting states and of their possible interplay in the layered iron pnictides, such measurements are crucial to gain more insight on the specific formation mechanism of the superconducting ground state. Finally, by detecting local field distributions, the $\mu$SR technique is used to test the nature of superconducting states by studying the temperature and field dependence of the density of superconducting carriers $n_s$, directly obtained through the determination of the London magnetic penetration depth $\lambda$ in the vortex phase.

\section{Principle of the $\mu$SR technique}
In $\mu$SR experiments, 100\% polarized muons are implanted in the solid sample. After a deceleration within
100~ps, which is too rapid to allow any significant loss of polarization and
sufficiently rapid for the $\mu$SR time window, the muon comes at rest at interstitial lattice sites and decays after an average lifetime of $\tau_{\mu} = 2.197~\mu$s, emitting a positron ($e^+$) and two neutrinos.

The parity violation in the weak interaction is reflected by an anisotropic distribution of
the positron emission with respect to the muon-spin direction at the
decay time. The positron emission probability is given by
\begin{equation}
 W(\theta)d\theta \propto (1 + A\cos\theta)d\theta~,
\end{equation}
where $\theta$ is the angle between the positron trajectory and the
muon-spin direction at the moment of the decay. 
This anisotropic positron emission constitutes the basics for
the $\mu$SR technique. By measuring the positron distribution, the original muon-spin direction can be determined. 
The asymmetry of $W(\theta)$ is given by $A = aP_{\mu}(0)$, where $P_{\mu}(0) =
|\mbox{\bf P}_{\mu}(0)|$ is the beam polarization, which is of the order of $\sim$1, and
$a$ is an intrinsic asymmetry parameter determined by the weak decay mechanism.
If all emitted positrons are detected with the same efficiency irrespective of
their energy, an average of $\langle a \rangle = \frac{1}{3}$ is obtained. 
Practically, during $\mu$SR measurements, only values up to $A \simeq 0.25$ can be obtained.

If the implanted muons are subject to magnetic interactions, their
polarization becomes time dependent [$\mbox{\bf P}_{\mu}(t)$]. 
The time evolution of $\mbox{\bf P}_{\mu}(t)$ can be deduced by measuring 
the positron distribution as a function of time. In the standard $\mu$SR technique, 
repeated measurements are made of the time interval between the $\mu^+$-implantation into the sample and the
detection of the emitted positron in a particular direction [say, in the
direction of the initial polarization $\mbox{\bf P}_{\mu}(0)$]. 
The time histogram of the collected intervals adopts therefore the form
\begin{equation}
 \label{equation_general}
  N_{e^+}(t) = B_g + N_0
  e^{-\frac{\displaystyle t}{\displaystyle\tau_{\mu}}}{\Big [}1 +
  A\frac{P_{\mu}(t)}{P_{\mu}(0)}{\Big ]}~,
\end{equation}
where $B_g$ is a time-independent background, $N_0$ is a normalization constant
and the exponential accounts for the decay of the $\mu^+$. $P_{\mu}(t)$ is
defined as the projection of $\mbox{\bf P}_{\mu}(t)$ along
the direction of the initial polarization, i.e. $P_{\mu}(t) = \mbox{\bf
P}_{\mu}(t) \cdot \mbox{\bf P}_{\mu}(0)/P_{\mu}(0) = G(t)P_{\mu}(0)$
where $G(t)$  reflects the normalized $\mu^+$-spin auto-correlation function
\begin{equation}
 G(t) = \frac{\displaystyle\langle \mbox{\bf S}(t)\cdot
 \mbox{\bf S}(0)\rangle}{\displaystyle S(0)^2}~,
\end{equation}
which depends on the distribution, average value and time evolution of the
internal fields and therefore contains all the physics of the magnetic
interactions of the $\mu^+$ inside the sample. Equation
(\ref{equation_general}) can be written as
\begin{equation}
 N_{e^+}(t) = B_g + N_0
 e^{-\frac{\displaystyle t}{\displaystyle\tau_{\mu}}}{\Big [}1 +
 A G(t){\Big ]}~.
\end{equation}
The term $AG(t)$ is often called the $\mu$SR signal and the envelope of
$G(t)$ is known as the $\mu^+$-depolarization function. 

Since the $\mu^+$ are uniformly implanted in the sample, the coexistence of
different domains, characterized by different types of ground states, will also
be detected by the presence of different components with distinct functions
$G_i(t)$. The amplitude ($A_i$) of each components will readily be
a measure of the volume fractions associated with different domains, 
with the condition that they sum up to the known maximal achievable asymmetry, i.e.:
\begin{equation}
 \sum_i A_i = A~.
\end{equation}
A potential difficulty is that in a particular crystallographic structure, 
more than one stopping sites can occur. They will also be identified by a 
$\mu$SR signal with different components,
i.e. with different spin auto-correlation functions $G_i(t)$. In this case 
the respective amplitudes ($A_i$) will reflect the probabilities to 
stop the muon at a given site. Therefore, great care has to be taken to 
differentiate this case from the occurrence of different ground states in the sample. 

If a static local
magnetic field ($\mbox{\bf B}_{\mu}$) is present at the $\mu^+$-site, $G(t)$ is given by
\begin{equation}
 \label{equation_timeevolpol}
  G(t) = \int f(\mbox{\bf B}_{\mu}){\Big [}\cos^2\theta +
  \sin^2\theta\cos(\omega_{\mu}t){\Big ]}d\mbox{\bf B}_{\mu}~,
\end{equation}
where $f(\mbox{\bf B}_{\mu})$ is the magnetic field distribution function,
$\theta$ is the angle between the local field and $\mbox{\bf P}_{\mu}(0)$. 
Hence the presence of internal magnetic fields in a sample will be revealed 
by the occurrence of frequencies $\nu_{\mu}$ in the $\mu$SR signal, which will be a direct measure of the internal fields as  
$\nu_{\mu} = \omega_{\mu}/(2 \pi) = \gamma_{\mu}/(2 \pi) B_{\mu}$, where 
$B_{\mu} = |\mbox{\bf B}_{\mu}|$ and $\gamma_{\mu}/(2 \pi) = 
135.53879 (\pm 0.7~\mbox{ppm})$~MHz/T is the gyromagnetic ratio of the muon. 
In practice, internal fields 
resulting either from a magnetic state (creating spontaneous frequencies) or from externally applied fields 
will have some distribution around an average value 
creating a depolarizing envelope of the oscillatory $\mu$SR signal, due to the dephasing of the muon ensemble. 

Such depolarization is exploited to determine the London penetration depth in a type II superconductor by performing measurements in the vortex phase created by applying a high transverse external magnetic field. From the observed depolarization rate $\sigma$, usually of Gaussian character, the penetration depth can be obtained recalling that in the clean limit:
\begin{equation}
\label{sup_density}
\sigma \propto \lambda^{-2} \propto \frac{n_s}{m^*}~,
\end{equation}
where $m^*$ is the effective mass of the carriers (see for example \cite{brandt}). By recording the temperature and field dependence of $\sigma$ and hence consequently $n_s$, indication about the likely occurrence of unconventional superconducting states can be pinpointed.
\section{Characterizing the magnetic state of mother compounds}
Early $\mu$SR studies have been performed to characterize the magnetic state of the first discovered family of FeAs layered superconductors, the so-called ``1111'' family with REFeAsO where RE represents a rare-earth ion. Studies have been reported on systems with RE = La \cite{klauss, carlo}, Nd \cite{aczel} and  Sm \cite{drew}. The compound LaFeAsO has at an early stage attracted interest as no complication of rare-earth magnetic ordering is at play. Both reported measurements indicate the presence of two spontaneous frequencies for temperatures below $T_{\rm N} = 138$~K. The temperature dependence of the well-defined frequencies indicates that the magnetic transition is of second-order and that the saturation values for $T \rightarrow 0$ are 23 and 3~MHz. The well-defined frequencies point to the presence of a commensurate static magnetic order and the amplitude of the observed magnetic $\mu$SR signal implies that 100\% of the sample volume orders at $T_{\rm N}$. Such findings are on-line with the reported magnetic structure obtained by neutron diffraction \cite{cruz}. The presence of different components points to the occurrence of two distinct muon stopping site, with the high frequency associated with muons stopping near the Fe magnetic moments in the FeAs layers, whereas the lowest frequency points to the presence of a more remote muon stopping site, most likely near the LaO layers. By comparing directly the $\mu$SR and M\"ossbauer data Klauss {\it et al.} \cite{klauss} determine a strongly reduced iron ordered moment of 0.25(5)$\mu_{\rm B}$. 

\begin{figure}
\includegraphics[width=8.7cm]{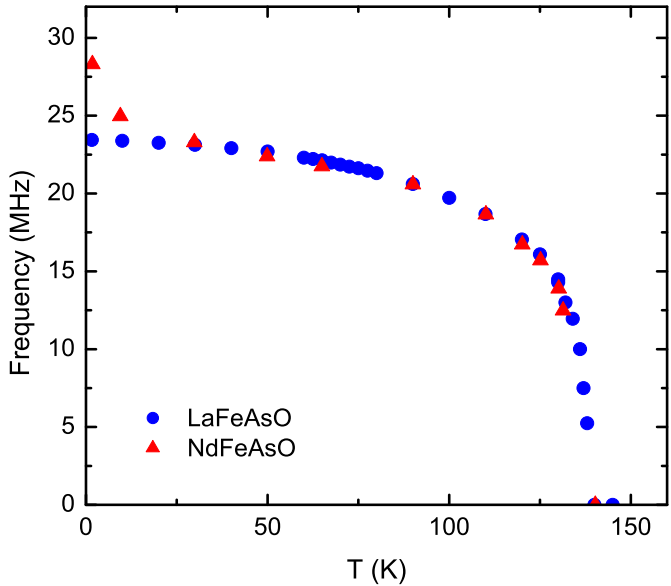}
\caption{Temperature dependence of the highest spontaneous $\mu$SR frequencies recorded for two systems of the ``1111'' family, namely LaFeAsO  \cite{klauss} and NdFeAsO \cite{aczel}. Note the change of the frequency value at low temperature for NdFeAsO reflecting the ordering of the rare-earth moments.}
\label{frequency_temp}
\end{figure}
The complication provided by the ordering of the rare-earth moments at low temperature is exemplified by the data reported by Aczel {\it et al.} \cite{aczel} on NdFeAsO (see Fig.~\ref{frequency_temp}). In addition to have a N\' eel temperature comparable to the one of LaFeAsO, the main frequency signal exhibits above about 5~K a very similar temperature dependence and absolute value to the ones observed in LaFeAsO. This indicates, in addition of suggesting analogous muon stopping sites, that the ordered Fe-moments for these two systems are of the same size for temperature regions unaffected by the rare-earth ions ordering. The magnetic ordering of the Nd sublattice is reflected by a jump of the frequency below about 5~K. 
The robustness of the ordering within the FeAs layers is also detectable by $\mu$SR on the Sm-based mother compound SmFeAsO \cite{drew}, where a similar spontaneous muon-spin precession frequency of 23.6~MHz is observed for the iron-moments ordering. Here also, a clear signature of the rare-earth ordering is observed at temperatures below 5~K. However, the $\mu$SR signal at very low temperatures exhibits the occurrence of additional frequencies \cite{drew}, pointing therefore to a different magnetic ordering of the Sm-sublattice compared with the case of NdFeAsO, where a mere shift of the main frequency was observed.

\begin{figure}[t]
\includegraphics[width=8.7cm]{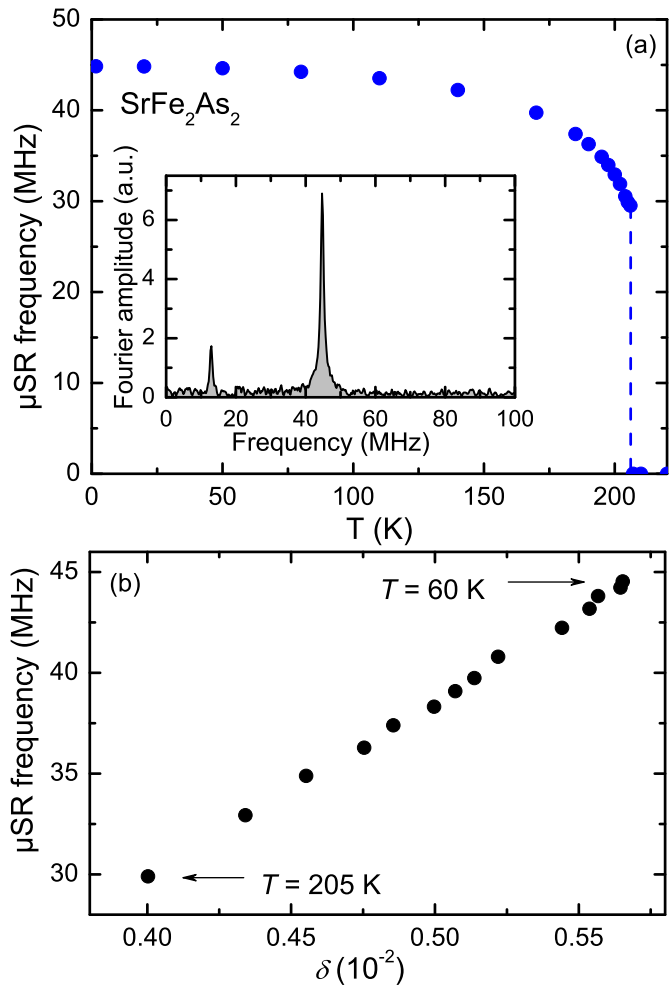}
\caption{(a) Temperature dependence of the highest muon-spin precession frequency recorded in SrFe$_2$As$_2$. Note the clear first-order transition at 205~K. The insert represents a Fourier transform of the $\mu$SR signal recorded at 1.6~K, where the high and low frequencies are clearly visible. (b) Observed scaling between the $\mu$SR frequency and the orthorhombic distortion, with the temperature as implicit parameter. Both panels are adapted from Ref. \cite{jesche}.}
\label{122}
\end{figure}

For all the pristine ``1111'' compounds investigated by $\mu$SR the formation of the magnetic state is confirmed to occur in a second-order transition and well separated from the temperature $T_S\simeq 160$~K where a structural transition from the high-temperature tetragonal (space group $P4/nmm$) phase to a orthorhombic  (space group $Cmma$) phase is observed \cite{nomura}.
A drastically different picture is obtained when investigating by $\mu$SR the mother compounds of the so-called ``122'' family. The ternary oxygen-free iron-based arsenide XFe$_2$As$_2$ (where X = Ca, Sr, Ba) systems crystallize in a high-temperature tetragonal structure of the ThCr$_2$Si$_2$-type (space group $I4/mmm$). Such structure contains almost identical layers of FeAs edge-sharing tetrahedra. Instead of LaO layers as in the ``1111'' family, these layers of tetrahedra are separated by ions of type X along the $c$-axis. However, the ``122'' family exhibits striking similarities with the ``1111'' series in the sense that a rather comparable tetragonal to orthorhombic (space group $Fmmm$) structural transition is observed upon lowering the temperature and that a magnetic ground state appears. Such magnetic order is characterized by a columnar antiferromagnetic ordering of the Fe moments with a propagation vector (1,0,1). On the contrary of what has been observed for the REFeAsO series, for the XFs$_2$As$_2$ systems Rotter {\it et al.} \cite{rotter} convincingly established that the structural and magnetic transition transitions occur simultaneously suggesting that the magnetic and orthorhombic order parameters are strongly coupled. However, quite some controversy still exists on the order of the  phases transitions, which was first characterized to be of second-order and later reported to be of the first-order \cite{rotter,huang,krellner}. As the space group $Fmmm$ of the low-temperature orthorhombic phase is a subgroup of the space group of the high-temperature tetragonal phase $I4/mmm$, such classification could be compatible with the observed structural phase transition. 
Recently a combined study on the system SrFe$_2$As$_2$ \cite{jesche} using thermodynamic and transport measurements as well as x-rays and $\mu$SR clearly determined the first-order character of both structural and magnetic transitions. The $\mu$SR spectra in SrFe$_2$As$_2$ resemble the ones reported for LaFeAsO with also two well-defined frequencies at 44 Mhz and 13~MHz, corresponding to 70\% and 30\% of the signal, respectively. The observed ratio between the highest frequencies in LaFeASO and SrFe$_2$As$_2$ is equal to the ratio of the hyperfine fields measured by M\"ossbauer indicating that in the ``122'' family, the muon stopping site near the FeAs layers is identical as the one occurring for the ``1111'' systems. In addition, the temperature dependence of the magnetic order parameter recorded by $\mu$SR (i.e. of the observed spontaneous frequencies) is basically identical to the one characterizing the orthorhombic distortion $\delta$ ($= a_O/(a_T\sqrt{2}) -1 = 1-b_O/(a_T\sqrt{2})$, where $a_T$, $a_O$ and $b_O$ are the  lattice parameters of basal plane for both structures) proving that the magnetic and orthorhombic phase are intimately coupled (see Fig.~\ref{122}). 
Such observation strongly suggests that the size of the orthorhombic distortion actually defines the size of the ordered iron moment, and therefore the muon-spin frequency, or vice versa. Interestingly, a scaling between the values of the orthorhombic distortion and of the muon-spin frequency, both extrapolated at $T = 0$, holds also for the LaFeAsO compound. A similar scaling factor of $\sim 92$~MHz (equivalent to almost exactly 1~$\mu_{\rm B}$ ordered iron moment) per percent of orthorhombic distortion is observed. Limited data is presently reported for the other compounds of the ``122'' family, i.e. X~=~Ca and Ba \cite{aczel,goko,uemura}. The observation of clear spontaneous $\mu$SR frequencies confirms the occurrence of well defined magnetic states. However, the few available points of the temperature dependence of the spontaneous $\mu$SR frequencies do not allow one to draw definite conclusions on the classification order of the magnetic transition. Here again, the coincidence between the disappearance of the spontaneous $\mu$SR frequencies and of the structural transitions clearly points to a close relation between both types of phenomena.

Another family of superconductors characterized by the presence of two-dimensional iron-based slabs has been also recently reported \cite{hsu}, i.e. the so-called ``011'' composed by the $\alpha$-FeSe$_{1-x}$ binary system. Such system is characterized by stacks of edge-sharing FeSe$_4$ tetrahedra, very similar to the one found in the oxypnictides. Interestingly, such system crystallizes in the PbO tetragonal structure (space group $P4/nmm$), but presents, at least for some anion concentration, below $T_S$ (of the order of 100~K) a structural transition into a orthorhombic phase (space group $Cmma$) showing remarkable similarities to that of the ``1111'' oxypnictides. Khasanov et al. \cite{khasanov} reported muon-spin measurements performed on a sample with nominal concentration FeSe$_{0.85}$. In the whole temperature region the zero-field data were found to be well described by the single-exponential decay function. This points either to the existence of fast electronic fluctuations measurable within the $\mu$SR time window or to a static magnetic field distribution caused by diluted and randomly oriented magnetic moments. However, it was found that an externally applied longitudinal field allows one to quench the depolarization due to internal fields, proving therefore their static character. By comparing the $\mu$SR data with magnetization studies, Khasanov et al. conclude that the magnetism observed is, most probably, caused by traces of Fe impurities.

Finally, $\mu$SR studies of the magnetic properties of the oxygen-free compounds SrFeAsF and CaFeAsF have been reported \cite{baker,takeshita}. These compound can be classified as the parent compound of a newly discovered series of fluoropnictide superconductors, where the doping on the divalent metal site may lead to superconductivity, as for example Sr$_{0.5}$Sm$_{0.5}$FeAsF which superconducts below $T_c = 56$~K \cite{wu}. 
Such family represents a variation of the ``1111'' family, where trivalent cation and oxygen are respectively replaced by divalent alkaline earth metal (Ca or Sr) and fluorine. The crystal structure is comparable to the ``1111'' oxypnictides with similar FeAs layers (ZrCuSiAs-type structure, space group $P4/nmm$), but divalent alkaline earth metal - fluoride layers replace the rare-earth - oxide layers. In this family, the fluoride ions are thought to provide weaker magnetic exchange pathways between the FeAs layers than for the above described ``1111'' oxypnictides or ``122'' families.  Again similarly to the oxypnictides, one observes for the parent compounds a structural transition to a low-temperature orthorhombic structure ($Cmma$) below ($T_S \simeq 175$~K for SrFeAsF \cite{tegel} and $T_S \simeq 134$~K for CaFeAsF \cite{xiao}). 

Concerning SrFeAsF, and based on M\"ossbauer studies \cite{tegel}, it was speculated that the structural transition was accompanied by a  concomitant magnetic transition. However, Baker et al. \cite{baker} clearly demonstrate by $\mu$SR that SrFeAsF possesses an even greater separation between the structural and magnetic ordering temperatures ($T_S-T_{\rm N} \simeq 50$~K) than in the ``1111'' oxypnictide family and that the magnetic transition is of second-order. Interestingly, the observed spontaneous frequencies observed below $T_{\rm N}$ are slightly lower that in LaFeAsO but in similar proportion, suggesting that the magnetic structure and ordered moments are comparable to LaFeAsO, and also points to related muon stopping sites. A difference with the oxypnictides is furnished by the temperature dependence of the frequencies, which can also be fitted by a conventional power law $\nu_i(T)=\nu_{i,0}(T)[1-(T/T_{\rm N})^{\alpha}]^{\beta}$ but with parameter 
$\beta = 0.22(3)$ representing a slightly sharper phase transition suggesting a more two-dimensional magnetic interactions than in oxypnictide compounds. This is consistent with the increased separation $T_S-T_{\rm N}$ and with the expectation that the interplanar exchange mediated by a fluoride layer is weaker than that mediated by an oxide layer in the ``1111'' family.

For the CaFeAsF compound, Takeshita et al. \cite{takeshita} observed a second order magnetic transition around 120~K, below which a single spontaneous $\mu$SR frequency appears and which saturates at about 25~´MHz. Hence, the magnitude of the internal field is again rather close to the ones observed for the oxypnictides ``1111'' family. 

To complete the picture, a careful study of the possible impurity phases (namely, FeAs and FeAs$_2$) in pnictides compounds containing FeAs layers has been reported by Baker et al. \cite{baker2}. FeAs is known to exhibit an helimagnetic order below $T_{\rm N}=77$~K \cite{selte}, whereas FeAs$_2$ has previously been shown not to order down to 5~K \cite{yuzuri}. The spontaneous $\mu$SR frequencies observed in FeAs for $T\rightarrow 0$~K are both strongly damped and have values of $\sim$23~MHz (corresponding to 70\% of the signal) and $\sim$38~MHz for the remaining signal amplitude. Above $T_{\rm N}$ the $\mu$SR signal is weakly damped as expected in the paramagnetic phase. The 23~MHz signal observed for FeAs is similar to that in the ``1111'' family and in the SrFeAsF compound  (see above). However, the ``1111'' compounds have no higher frequency signal, and a much lower damping rate. Also, the 23~MHz signal in the ``1111'' compounds persists all the way up to $T_{\rm N}\simeq 140$~K, twice that of FeAs. Consequently, Baker et al. concluded that the local environments for implanted muons are relatively similar in FeAs and the ``1111'' family, but FeAs impurities cannot produce the observed signal. A similar conclusion can be drawn for SrFeAsF, where the observed $\sim$23~MHz signal persists up to $T_{\rm N}\simeq 120$~K and therefore is not related to FeAs impurities. The $\mu$SR studies of the impurity phase FeAs$_2$ confirm the absence of static ordering, with the observation of a weak depolarization most probably reflecting the effect of the atomic-moments fluctuations within the $\mu$SR time-window. 

\section{Evolution of the magnetic state upon doping and its interplay with superconductivity}
Like the cuprate high-$T_c$ superconductors, the superconducting state in the iron-based layered systems is reached upon charge doping the magnetic parent compound. Such doping can be achieved through electron doping by creating oxygen deficit, substituting oxygen by fluorine in the RE-oxygen layers, or even replacing iron by e.g. cobalt in the FeAs layers. Another route is hole doping, achieved by a partial substitution of the rare-earth ions (e.g. La by Sr). As in the cuprates, such doping leads to a weakening of the magnetic state as the superconducting state emerges raising the question of the nature of the interplay between both ground states. As $\mu$SR studies have provided unique information on the evolution of the magnetism upon doping in cuprates \cite{niedermayer}, it is of no surprise that early and extensive $\mu$SR investigations have also been devoted to the search of possible generic features common to cuprates and iron-based layered systems, which could shade more light on the understanding of the mechanisms of high-$T_c$ superconductivity.

\begin{figure}[tb]
\includegraphics[width=8.7cm]{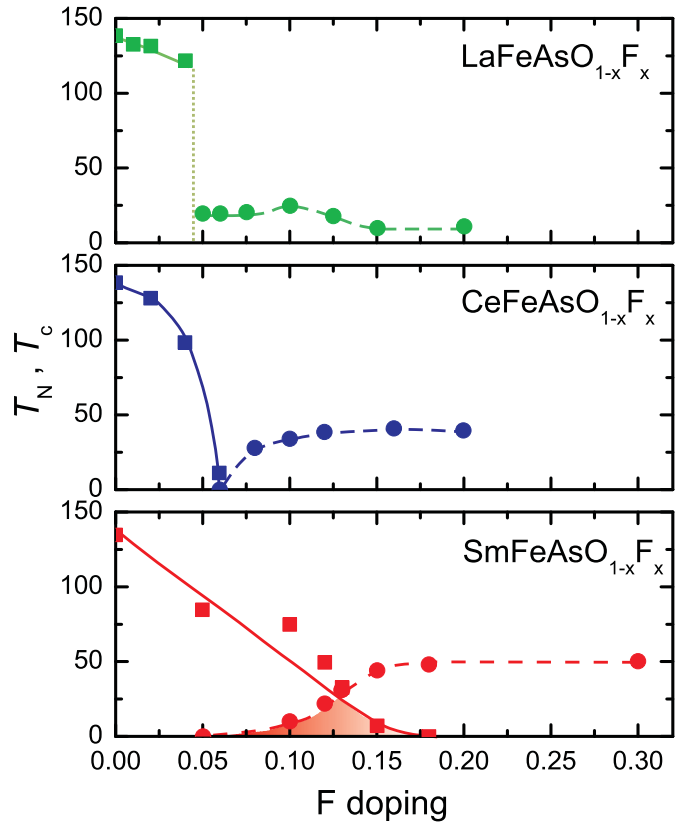}
\caption{Phase diagrams  as a function of fluorine concentration for different systems of the ``1111'' family REFeAsO$_{1-x}$F$_x$. The data with RE = La \cite{luetkens} and RE = Sm \cite{drew} are obtained by $\mu$SR, whereas the data on RE = Ce were obtained by neutron diffraction \cite{zhao}. Dashed lines and circles: $T_{\rm c}$; Solid lines and squares: $T_{\rm N}$. Note the first-order-like transition for the RE = La system, the apparent quantum critical point for RE = Ce and the coexistence for the RE = Sm system (shaded area).}
\label{phase_diagrams}
\end{figure}

Much interest has been first dedicated to the study of the lanthanum member of the ``1111'' family, as the lack of a $4f$-electrons contribution allows one to gain a clearer view on the evolution of the Fe-moments magnetism. First $\mu$SR data on doped but still magnetic LaFeAsO$_{1-x}$F$_x$ were reported by Carlo et al. \cite{carlo}. By doping with 3\% fluorine, the ZF-$\mu$SR spectra exhibit a change from the simple cosine signals observed in undoped LaFeAsO towards a Bessel function line shape. Such a change hints to a crossover from a commensurate magnetic structure (CMS) towards an incommensurate one (IMS). Whereas, for CMS sharp distributions are present due to the constant phase between the muon-stopping site and the spin modulation, for IMS the field distribution seen by the muon represents a sinusoidal distribution of internal fields and is given by \cite{major}:
\begin{equation}
\label{equation_nims}
f_{{\rm IMS}}(B) = \left\{\begin{array}{ll}\frac{\displaystyle
2}{\displaystyle
\pi}\frac{\displaystyle 1}{\displaystyle \sqrt{B^2-B_{\rm max}^2}}
& \mbox{for}~B < B_{{\rm max}}\\
~\\
0 & \mbox{otherwise.} \end{array}\right.
\end{equation}
By combining Eq.~\ref{equation_nims} and Eq.~\ref{equation_timeevolpol}, one can easily calculate that the time evolution of the muon polarization is given by the $J_0$ Bessel function. Carlo et al. \cite{carlo} associate such change to the one observed in the cuprate La$_2$CuO$_4$ at the 1/8 doping when the formation of spin stripes occurs \cite{savici}. 
In the cuprates, such stripe order involves a spatial segregation of doped holes. This results into a persistence of antiferromagnetic hole-poor regions. An alternative route to explain the incommensurability of the magnetic structure upon doping is to consider that the increase of electron doping changes slightly the topology of the different Fermi surfaces thereby modifying the length of the antiferromagnetic nesting vector responsible for the magnetic order \cite{maier}. Such explanation was invoked by Luetkens et al. \cite{luetkens} as they reported the observation of a similar Bessel function in a LaFeAsO$_{0.96}$F$_{0.04}$ sample. Such view point appears somewhat corroborated by the observation, at very low doping, of a smooth evolution of both the magnetic order temperature transition and the ordered value of the magnetic moment \cite{luetkens}. The rather systematic study of the evolution of the magnetism upon doping in the LaFeAsO$_{1-x}$F$_{x}$ series presented by Luetkens and coworkers \cite{luetkens} also clearly demonstrates the total absence of any coexistence between magnetism and superconductivity. Hence, for a fluorine doping $x$ between 0.04 and 0.05, a sharp first-order-like transition from spin-density wave magnetic order towards superconductivity is observed. The abrupt change of the ground state is visible in the transition temperatures $T_{\rm N}$ and $T_c$ (see Fig.~\ref{phase_diagrams}), as well as on the magnetic order parameter. It this work the abrupt change was not only mapped out by $\mu$SR but also by M\"ossbauer spectroscopy. Interestingly, such first-order-like transition is also observed on the orthorhombic distortion which also disappears at the same doping, as demonstrated by x-ray diffraction on the same samples. This confirms the close correlation between the structural distortion and the magnetic ground state properties. 
Moreover, the steplike behavior of the superconducting order parameter, between $x = 0.04$ and 0.05 suggests that
the key element for superconductivity is the suppression of the orthorhombic distortion, and therefore of the static magnetic order, rather than the moderate increase of the charge carrier density by the fluorine doping. 
More importantly, the study of Luetkens and coworkers rules out the occurrence of a magnetic quantum critical point, at least in La-based ``1111'' systems.

Note that in a recent neutron scattering study of the doping dependence of the orthorhombic distortion and the magnetic order in CeFeAsO$_{1-x}$F$_x$ \cite{zhao} a gradual suppression of $T_{\rm N}$ as a function of the fluorine doping within the orthorhombic phase was observed, pointing to the presence of a quantum critical point (see Fig.~\ref{phase_diagrams}). However, subsequent $\mu$SR experiments on the very same sample with the magnetically ordered composition close to the phase boundary ($x=0.06$) reveals an increased magnetic ordering temperature \cite{dai}. A similar systematic $\mu$SR study as the one performed for the La-based system is therefore urgently needed for the Ce-based system.

Finally, in the Sm-based system of the ``1111'' family, $\mu$SR studies appear to reveal a third type of behaviour. Hence, a microscopic coexistence of bulk magnetic order and superconductivity \cite{drew}, connected with a gradual suppression of N\'eel transition temperature and the persistence of slow magnetic fluctuations, was observed in the doping range $0.10 < x < 0.17$ (see Fig.~\ref{phase_diagrams}). It must be noticed that for the high-$T_c$ cuprates a similar picture occurs as the onset of superconductivity gives rise to a suppression of the local magnetic order and the emergence of slow magnetic fluctuations, persisting well deeply into the superconducting state. In view of their results, Drew et al. \cite{drew} noticed that in the Sm-based system where static magnetism coexists with superconductivity and slow magnetic fluctuations persist into the superconducting state \cite{drew2,khasanov2}, the maximal $T_c$ is almost double that of the LaFeAsO$_{1-x}$F$_{x}$ systems, where spin fluctuations are hardly detectable in the superconducting samples by $\mu$SR. As mentioned above, slow spin fluctuations emerging at the boundaries of a magnetic state can induce and/or enhance the Cooper pairs formation mechanism usually connected with a unconventional superconducting order parameter \cite{kato}. In view of the results reported by Khasanov et al. \cite{khasanov2} the question is raised about the origin of the observed slow fluctuations for high-doping.
Hence, one observes that the temperature dependence of the fluctuation rate is characterized by an activation energy, most probably reflecting the thermal population of samarium crystal-field levels. This is supported by the value of the fitted activation energy of $\sim$23~meV, which is in good agreement with the first excited crystal field level of the Sm$^{3+}$ ions
deduced from the observation of Schottky anomalies in the specific heat \cite{baker3}.
Luetkens et al. \cite{luetkens} suggested that the origin for the difference between the Sm and Ce-based systems to the LaFeAsO$_{1-x}$F$_{x}$ ones might be related to a strong electronic coupling to the rare-earth magnetic moments or possibly to a different level of local structural distortions at the $x$-dependent crossover from the low temperature orthorhombic lattice structure to the tetragonal phase.

The connection between the presence of rare-earth ions in the system and the observation of coexistence between magnetism and superconductivity does not hold for the fluoropnictide system Ca(Fe$_{1-x}$Co$_x$)AsFe. As shown by Takeshita et al. \cite{takeshita}, the substitution of iron by cobalt leads to a weakening of the magnetic state and a mesoscopic coexistence between both ground states. However, one also observes that upon doping the magnetic order is no more characterized by well defined frequencies, but rather by a fast depolarization, most probably reflecting a large disorder of the magnetic structure connected with the iron ions substitution. 

Concerning the ``122'' family (XFe$_2$As$_2$, where X = Ca, Sr, Ba -- see above), several $\mu$SR measurements have been devoted to the study of the hole doping, which results into a weakening of the magnetic state and the occurrence of superconductivity. Such doping can been obtained, for example, through the substitution of divalent cations (X$^{2+}$) through potassium as in Ba$_{1-x}$K$_x$Fe$_2$As$_2$. For this latter system, quite some divergence appears when comparing the available $\mu$SR studies. In line with transport, x-ray and neutron diffraction studies (see for example \cite{chen}), some $\mu$SR studies \cite{aczel,goko,park} report that the magnetic order and the superconductivity overlap in the intermediate composition range. However, all $\mu$SR data appear to exclude any microscopic coexistence between both ground states, which are therefore spatially separated, with a magnetic order exhibiting a rather large correlation length $> 100$\AA.
Early measurements by Aczel et al. \cite{aczel} suggest that up to $\sim$80\% of the volume of a $x=0.45$ single-crystalline sample maintain a magnetic ground state and that superconductivity is limited to the microscopically separated non-magnetic fraction. As for the ``1111'' family a shift from a commensurate to an incommensurate magnetic structure is evidenced by the occurrence of a Bessel rather than a cosine function to characterize the $\mu$SR spectra (see Eq.~\ref{equation_nims}). Goko et al. \cite{goko} report that for a $x=0.5$ sample, i.e. possessing a nominal optimal doping, the magnetic fraction is still $\sim$50\% at low temperature, i.e. well below the superconducting transition temperature. 
Finally, Park et al. \cite{park} state that in a $x=0.41$ sample, the superconducting fraction is about 1/4 of the total volume, with $\sim$50\% still occupied by the usual magnetic state and the rest presenting a so far unknown ``disordered magnetic phase'' persisting up to room temperature. Such ``hidden'' order, which cannot be explained by impurities of Fe$_2$As phase, is tentatively ascribed to the presence of a weakly temperature-dependent density-wave-like order observed by ARPES technique in the very same sample \cite{zabolotnyy}. As for the previous measurements, one observes a gradual increase of the magnetic fraction below $T_{\rm N}$ rather than a sharp phase transition as in the undoped sample \cite{aczel,goko,uemura,jesche}. Note that for the last study, XRPD data indicate an homogeneous distribution of the dopant atoms and that the average potassium content was experimentally determined by looking at the functional dependency of the lattice parameters.

Such a presence of magnetic and superconducting ground states in the Ba$_{1-x}$K$_x$Fe$_2$As$_2$ systems is at first sight similar to the situation reported for the cuprates high-$T_c$ in the underdoped regime. However, for the cuprates both ground states appear to coexist at the microscopic level (see for example \cite{niedermayer}), whereas a mesoscopic phase separation is observed in the iron-layered superconductors.

To add more complexity to the picture of the doping-dependence of the magnetic state in the ``122'' family, we note that, very recently, Hiraishi et al. \cite{hiraishi} reported $\mu$SR data on a Ba$_{1-x}$K$_x$Fe$_2$As$_2$ polycrystalline sample with a nominal $x=0.4$ composition, where no trace of magnetism was found down to the lowest temperatures and with a superconducting volume fraction of 100\%. It remains to be seen whether such apparent divergence between the reported data is a result of a samples quality issue and/or of difference of doping between nominal and effective values.   

\section{On the nature of the superconducting state}
As said above, in a type-II superconductors, the absolute value and temperature dependence of the London magnetic penetration depth $\lambda$ can be determined from $\mu$SR transverse-field experiments by monitoring the depolarization of the $\mu$SR signal arising from the dephasing created by the flux-line lattice. The temperature dependence of the superconducting density $n_s$, which is directly related to the penetration depth (see Eq.~\ref{sup_density}), provides indications on the topology of the superconducting gap. Such technique has been widely used to characterized the superconducting state of cuprates (see for example \cite{sonier}) and has also lead to the so-called Uemura universal scaling (``Uemura plot'' \cite{uemura2}) between the superconducting transition $T_c$ and the superconducting carriers density. The BCS theory predicts that $T_c$ is related to the energy scale of attractive interaction, while only a weak and indirect dependence is expected on $n_s$. The observed scaling is taken as indication that the condensation in cuprates is related to a Bose-Einstein condensation mechanism of pre-formed pairs. In view of the apparent similarities between cuprates and iron-based pnictides, much interest has been focused on the characterization of the superconducting state by $\mu$SR on these latter systems.

Let us first discuss the specific information obtained by the temperature dependence of the penetration depth. For the ``1111'' family, the most complete picture has been obtained on the La-based systems \cite{luetkens2,luetkens}. Here again, the absence of $4f$ magnetic contribution at low temperatures provides a clearer answer than in rare-earth analogs. In anisotropic superconductors like ``1111'' family \cite{weyeneth} the muon depolarization rate can be converted into $\lambda_{ab}$, the in-plane magnetic penetration depth, and Eq.~\ref{sup_density} becomes \cite{brandt}:
\begin{equation}
\label{sup_density_2}
\lambda^{-2}_{ab}=  \frac{\sigma }{0.0355 \Phi_0 \gamma_{\mu}}\propto \frac{n_s}{m^*}~.
\end{equation}
Luetkens et al. \cite{luetkens2,luetkens} found that $\lambda^{-2}_{ab}$ follows, for $x$ between 0.05 and 0.10, a nearly temperature independent behavior below $T_c/3$ indicative of a low density of states in the superconducting gap. Whereas the temperature dependence could be compatible with a conventional BCS $s$-wave gap, the field dependence of $\lambda^{-2}_{ab}$ points to the occurrence of a dirty $d$-wave gap or a multi-gap superconducting state. This later possibility is backed, for example, by high-field measurements \cite{hunte} and recent NQR results \cite{kawasaki}. Moreover, the clear concave curvature of $\lambda^{-2}_{ab}$ for higher $x$-concentration near $T_c$ \cite{luetkens} appears also as a signature of multi-band effects (see for example \cite{khasanov3}). A similar conclusion was obtained by Takeshita et al. on a $x=0.06$ sample \cite{takeshita2}.

The presence of multi-gaps as a common feature among iron-based layered systems is also supported by the reported $\mu$SR measurements performed on the ``122'' Ba$_{1-x}$K$_x$Fe$_2$As$_2$. The clearest two-gap signature is furnished by the measurements of Khasanov et al. \cite{khasanov4}. On a single-crystal with $T_c = 32$~K, they report the experiments 
performed in an external magnetic fields applied in parallel and perpendicularly to the crystallographic $c$-axis, allowing them to separate the in-plane ($\lambda_{ab}$) and the out-of-plane ($\lambda_{c}$) components of penetration depth $\lambda$ and to reconstruct its anisotropy $\gamma_{\lambda} = \lambda_{c}/\lambda_{ab}$. Note that the anisotropy for the ``122'' family is about one order of magnitude lower than for ``1111'' systems. From the temperature dependence of $\lambda_{ab}^{-2}$, and the observation of a clear inflection point at $T=7$~K, they deduce a multi-gap behavior with zero-temperature values of the gaps estimated to be $\Delta_1 = 9$~meV and $\Delta_2 = 1.5$~meV. 
The temperature behavior of the anisotropy was found to increase upon decreasing temperature from $\gamma_{\lambda} = 1.1$ at $T_c$ to 1.9 at $T\simeq 1.7$~K. This resembles very much the situation in double-gap MgB$_2$ where the anisotropy of the penetration depth was found to be equal to the anisotropy of the critical fields at $T_c$, but has an opposite temperature
dependence. 

Other measurements on Ba$_{1-x}$K$_x$Fe$_2$As$_2$ systems were reported in Reference  \cite{aczel} and \cite{hiraishi}. Whereas the measurements by Aczel et al. \cite{aczel} on a $T_c \simeq 30$~K do not allow a reliable detailed study of the $T$-dependence of $n_s \propto \lambda^{-2}$, due to the strongly reduced superconducting volume fraction, they clearly point to the absence of points or nodes on the superconducting gap(s). A clearer picture was obtained on a $T_c \simeq 38$~K sample by Hiraishi et al. \cite{hiraishi}, where $n_s \propto \lambda^{-2}$ was found below $0.4T_c$ to be mostly independent of temperature, confirming the nodeless character of the superconducting gaps. Hence, Hiraishi et al. reported that the  temperature dependence of $n_s$ is perfectly reproduced by the conventional BCS model for $s-$wave paring, where the order parameter can be either a single-gap with $\Delta = 8.35(6)$~meV, or double-gap structure with $\Delta_1 = 12$~meV and $\Delta_2 = 6.8(3)$~meV. 

The observation of possible multi-gap features on the temperature dependence of $n_s$ is in-line with ARPES studies consistent with the fact that superconductivity occurs on complex Fermi surfaces consisting of many bands, that would give rise to a intricate temperature dependence of the density of superconducting carriers. In addition, the observed dependence of the gap values as a function of $T_c$ is fully compatible with different ARPES studies \cite{evtushinsky,ding}.

\begin{figure}[tb]
\includegraphics[width=8.7cm]{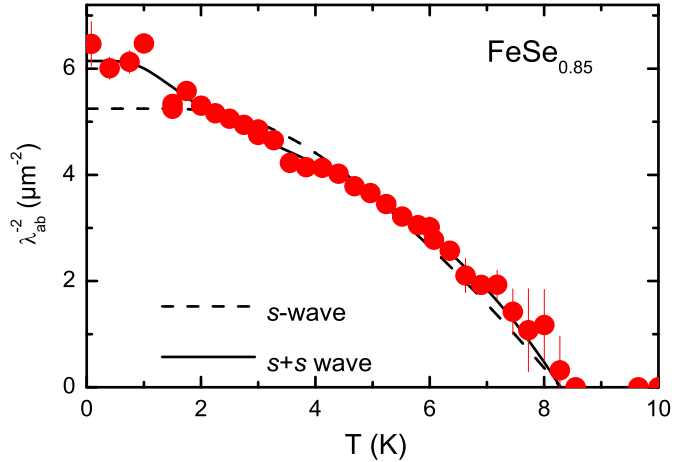}
\caption{Temperature dependence of $\lambda_{ab}^2$ obtained on FeSe$_{0.85}$ from the depolarization rate of the $\mu$SR signal. The curves represents fits assuming a single $s$-wave gap symmetry (dashed line) and a double $s$-wave gap (solid line). Adapted from Ref.~\cite{khasanov}.}
\label{FeSe}
\end{figure}

Concerning the temperature dependence of $n_s$ measured by $\mu$SR, and for completeness, one should also mentioned the measurements on the ``011'' family \cite{khasanov}, on the oxygen-free fluoropnictides \cite{takeshita} and on the LiFeAs (``111'') system \cite{pratt}. For the ``011'' system FeSe$_{0.85}$, clear evidence for nodeless superconductivity was reported. Here again, the curvature observed in $n_s(T)$ points to the occurrence of two gaps of $s$-wave character, or possibly to the presence of an anisotropic $s$-wave gap \cite{khasanov} (see Fig.~\ref{FeSe}). Measurements performed on CaFe$_{1-x}$Co$_{x}$AsF for $x = 0.075$ and $0.150$ show that also in the fluoropnictides the $T$-dependence of $n_s$ cannot be reproduced by a single $s$-wave BCS gap, but require an approach invoking at least a two-gap model \cite{takeshita}. Finally, for the ``111'' system LiFeAs, which is a new variant of iron-based layered superconductors without lanthanide-oxide layer, again $n_s(T)$ cannot be described by a simple $s$-wave BCS model as $n_s$ increases monotonically upon cooling below $T_c$ \cite{pratt}.

\begin{figure}
\includegraphics[width=8.7cm]{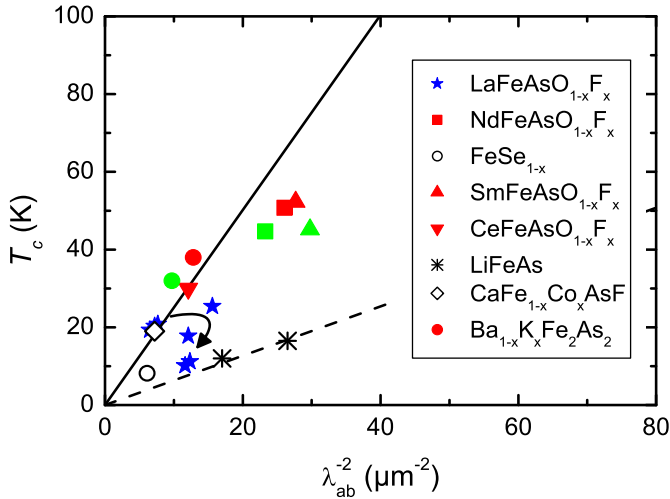}
\caption{Uemura plot for the iron-based layered systems with some of data obtained to date. The scalings observed for the high-$T_c$ cuprates is provided for comparison (solid line for hole doping; dashed line for electron doping). 
LaFeAsO$_{1-x}$F$_x$ data from Ref.~\cite{luetkens} -- Note the ``boomerang'' behavior upon increasing the fluorine doping (curved arrow); 
SmFeAsO$_{1-x}$F$_x$ data from Ref.~\cite{drew} (green) and \cite{khasanov2} (red); 
NdFeAsO$_{1-x}$F$_x$ taken from Ref.~\cite{carlo} (green) and \cite{khasanov2} (red); 
LiFeAs taken from Ref.~\cite{pratt}; 
CaFe$_{1-x}$Co$_x$AsF taken from Ref.~\cite{takeshita}; 
Ba$_{1-x}$K$_x$Fe$_2$As$_2$ taken from Ref.~\cite{goko} (green) and \cite{khasanov4} (red); 
FeSe$_{1-x}$ taken from Ref.~\cite{khasanov}.}
\label{uemura_plot}
\end{figure}

In Fig.~\ref{uemura_plot}, we report the $\mu$SR data available to date for iron-based layer systems on a ``Uemura'' plot. As discussed above, the extracted penetration-depth parameter $\lambda_{ab}^2$ is directly proportional to the supercarriers density $n_s$. Although no perfect single scaling is observed, one can tentatively extract the following trends. The ``1111'' and ``122'' families are found to follow rather closely the scaling observed for hole-doped cuprates. For the 
``111'' family, one seems to observe lower values than the trend line of the 
other systems. As pointed out by Pratt et al. \cite{pratt}, two lines of thoughts can obviously be invoked to explain such deviation. Either such systems possess supercarriers densities that are enhanced over the expected values on the basis of their $T_c$'s, or, alternatively, superconductivity is somehow suppress on those systems. Such reduction could be linked to crystal structure considerations. For all the families, the superexchange between Fe atoms through As is thought to provide the main contribution to the exchange interaction \cite{yildirim} and therefore one expect a dependence of the electronic properties on the angle between two legs of the tetrahedron formed by the As ions around the iron ions. Hence, one observe a clear decrease of this angle (and therefore shortening of the As-Fe distance) going from the ``1111'', through the ``122'' (presenting almost  regular As tetrahedra) and to ``111'' family. Roughly speaking, it is expected that a weaker Fe-As hybridization causes a decrease of the conduction bandwidth, accompanied by an increase in the density of states at the Fermi level $N(E_{\rm F})$ and the occurrence or strengthening of the superconducting state. Within a family, it is suggested that any deviation from the regular As tetrahedron is detrimental for superconductivity \cite{lee}. For example, for the LaFeAsO$_{1-x}$F$_x$, a so-called boomerang effect is observed for high fluorine concentration.

Although the similarities between the Uemura plots for cuprates and iron-based layer superconductors are apparently striking, it remains to be seen whether the underlaying mechanism to promote the superconducting state is common. Hence fundamental differences remain, as the possibility to induce superconductivity in the Fe layers by substituting Co or Ni ions directly into the Fe sites. This is in deep contrast with the case of cuprates, where the doping with Zn ions with one extra electron into Cu sites destroys superconductivity. Also, at least in the ``122'' family,  the antiferromagnetic order of the parent compound XFe$_2$As$_2$ can be destroyed by external pressure, which induces superconductivity without external doping. Such effect which is not present for cuprates stresses the role of the structural parameter as a key factor for the occurrence of high temperature superconductivity.

\section{Conclusions}
As for the cuprate systems, the $\mu$SR technique has very quickly provided valuable information on the ground state properties of these new high-$T_c$ superconductors. The specific advantages of the technique were exploited in particular to disentangle the interplay between magnetism and superconductivity, connected to ground states with almost degenerate condensation energies. In particular, and in almost all cases, one does not observe a microscopic coexistence of both phases, as in the cuprates, but rather a mesoscopic phase separation. However, the systems SmFeAsO$_{1-x}$F$_x$ seem to represent an exception as $\mu$SR data point to a microscopic coexistence of both phases. The close connection between the structural transition and magnetism has been clearly evidenced by $\mu$SR in particular on LaFeAsO$_{1-x}$F$_x$ systems. Moreover a scaling between $T_c$ and the superconducting carriers density, similar to the one observed on cuprates, appears also to occur in the iron-based layered pnictides. It remains however to be seen whether such scaling has a similar origin.

\end{document}